# Emergence of Self-Organized Amoeboid Movement in a Multi-Agent Approximation of *Physarum polycephalum*

Jeff Jones and Andrew Adamatzky

Centre for Unconventional Computing,
University of the West of England, Coldharbour Lane, Bristol BS16 1QY, UK.

jeff.jones@uwe.ac.uk, andrew.adamatzky@uwe.ac.uk

**Abstract:**

The giant single-celled slime mould *Physarum polycephalum* exhibits complex morphological adaptation and amoeboid movement as it forages for food and may be seen as a minimal example of complex robotic behaviour. Swarm computation has previously been used to explore how spatio-temporal complexity can emerge from, and be distributed within, simple component parts and their interactions. Using a particle based swarm approach we explore the question of how to generate collective amoeboid movement from simple non-oscillatory component parts in a model of *P. polycephalum*. The model collective behaves as a cohesive and deformable virtual material, approximating the local coupling within the plasmodium matrix. The collective generates *de-novo* and complex oscillatory patterns from simple local interactions. The origin of this motor behaviour is distributed within the collective rendering is morphologically adaptive, amenable to external influence, and robust to simulated environmental insult. We show how to gain external influence over the collective movement by simulated chemo-attraction (*pulling* towards nutrient stimuli) and simulated light irradiation hazards (*pushing* from stimuli). The amorphous and distributed properties of the collective are demonstrated by cleaving it into two independent entities and fusing two separate entities to form a single device, thus enabling it to traverse narrow, separate or tortuous paths. We conclude by summarising the contribution of the model to swarm based robotics and soft-bodied modular robotics and discuss the future potential of such *material* approaches to the field.

**Keywords:** *amoeboid movement, swarm intelligence, distributed computation, collective behaviour, Physarum polycephalum*

## Introduction:

Classical robotics approaches typically connect separate sensory, control, and locomotion systems, using complex parts with little redundancy. Nature inspired robotics takes inspiration from aggregate populations, collective transport, segmentation of component parts, and soft-bodied motion in living systems. Swarm approaches to nature inspired computing seek to elucidate the sensory mechanisms and individual interactions which generate the complexity patterning and movement seen at very different scales in natural systems including car traffic dynamics (Helbing, 2001), human walking patterns (Helbing et al., 1998), (Helbing et al., 2001), flocking and schooling (Reynolds, 1987), collective insect movement (Buhl et al., 2006), (Kube and Bonabeau, 2000), and bacterial patterning (Matsushita et al., 1999), (Ben-Jacob, 2003). In all these examples there is a population of entities in space, coupled by sensory information. The different sensory coupling mechanisms between individuals in these apparently disparate systems has also been abstracted at a minimal level of self-propelled particles to find common mechanisms (Vicsek et al., 1995).

Swarm approaches typically approximate collective movement at the *population* level. Distributed generation and control of movement also occurs at the individual level (Kennedy et al., 2001). Specific examples include flagellated movement of bacteria (Zhang et al., 2009), eukaryotes (Dreyfus et al., 2005), ciliated transport (Suh et al., 2000), peristaltic propulsion (Trimmer et al., 2006), (Saga and Nakamura, 2004), and amoeboid movement (Yokoi et al., 2003), (Umedachi et al., 2009). Many organisms make use of segmentation, employing identical or similar sub-units to form their body plan and generate structure and movement. Such modularity is attractive to computer science and robotics but presents challenges in terms of how to configure, communicate between, and control the devices. Population based approaches have also proved useful in tackling some of these issues. Internal propagation of simulated hormone signals was used to control gait patterns in simulated caterpillar-like modular robots (Salemi et al., 2001). Reconfiguration stimulated by crystalline mechanical transformation within a population of automata was demonstrated in (Rus and Vona, 2000) and future possibilities (and difficulties) in communicating between entities at larger scales with massive populations were discussed in (Goldstein et al., 2009). The exploitation of external environmental attractants ("seeds" and "scents") was used to effect the reconfiguration of a simulated modular robotic system composed of simple finite state machines for the creation of specific manipulation structures (Bojinov et al., 2002), and pre-desired shapes (Stoy, 2006).

Progress in biologically inspired robotics has also been made by considering the use of even simpler structures which straddle the boundary of non-living physical materials and living organisms including those acting as biological fibres and membranes (Zhang, 2003), lipid self assembly in terms of networks (Lobovkina et al., 2008), pseudopodium-like membrane extension (Lobovkina et al., 2009) and even those exhibiting simple chemotaxis responses (Lagzi et al.). Some engineering and biological insights have already been gained by studying the structure and function of what might be termed 'semi-biological' materials and the complex behaviour seen in such minimal examples raises questions about the lower bounds necessary for the emergence of apparently intelligent behaviour.

An ideal hypothetical candidate for a biological machine would be an organism which is capable of the complex sensory integration, movement and adaptation of a living organism, yet which is also composed of a relatively simple material that is amenable to simple understanding and control of its properties. We suggest that the myxomycete organism, the true slime mould *Physarum polycephalum*, is a suitable candidate organism which meets both criteria; i.e. it is a complex organism, but which is composed of relatively simple materials. A giant single-celled organism, *P. polycephalum* is an attractive biological candidate medium for emergent motive force because the basic physical mechanism during the plasmodium stage of its life cycle is a self-organized system of oscillatory contractile activity which is used in the pumping and distribution of nutrients within its internal transport network. The organism is remarkable in that the control of the oscillatory behaviour is distributed throughout the almost homogeneous medium and is highly redundant, having no critical or unique components.

The plasmodium is amorphous in shape and ranges from the microscopic scale to up to many square metres in size. It is a single cell syncytium formed by repeated nuclear division, comprised of a sponge-like actomyosin complex co-occurring in two physical phases. The gel phase is a dense matrix subject to spontaneous contraction and relaxation, under the influence of changing concentrations of intracellular chemicals. The protoplasmic sol phase is transported through the plasmodium by the force generated by the oscillatory contractions within the gel matrix. Protoplasmic flux, and thus the behaviour of the organism, is affected by changes in pressure, temperature, space availability, chemoattractant stimuli and illumination (Carlile, 1970; Durham and Ridgway, 1976; Kishimoto, 1958; Nakagaki et al., 1996; Nakagaki et al., 2000b; Takamatsu et al., 2000; Ueda et al., 1975). The *P. polycephalum* plasmodium can thus be regarded as a complex functional material capable of both sensory and motor behaviour. Indeed *P. polycephalum* has been described as a membrane bound reaction-diffusion system in reference to both the complex interactions within the plasmodium and the rich computational potential afforded by its material properties (Adamatzky et al., 2008). The study of the computational potential of the *P. polycephalum* plasmodium was initiated by Nakagaki et al. (Nakagaki et al., 2000a) who found that the plasmodium could solve simple maze puzzles. This research has been extended and the plasmodium has demonstrated its performance in, for example, path planning and plane division problems (Shirakawa and Gunji, 2010), (Shirakawa et al., 2009), spanning trees and proximity graphs (Adamatzky, 2007),(Adamatzky, 2008), simple memory effects (Saigusa et al., 2008), the implementation of individual logic gates (Tsuda et al., 2004) and *P. polycephalum* inspired models of simple adding circuits (Jones and Adamatzky, 2010).

From a robotics perspective it was shown that by its adaptation to changing conditions within its environment, the plasmodium may be considered as a prototype micro-mechanical manipulation system, capable of simple and programmable robotic actions including the manipulation (pushing and pulling) of small scale objects (Adamatzky and Jones, 2008), transport and mixing of substances (Adamatzky, 2010c) and as a guidance mechanism in a biological—mechanical hybrid approach where the response of the plasmodium to light irradiation was used to provide feedback control to a robotic system (Tsuda et al., 2007). A *P. polycephalum* inspired approach to amoeboid robotics was demonstrated by Umedachi et. al. (Umedachi et al., 2009) in which an external ring of coupled oscillators, each connected to passive and tuneable springs was coupled to a fluid filled inner bladder. The compression of the peripheral springs mimicked the gel contractile phase and the flux of sol within the plasmodium was approximated by the coupled transmission of water pressure to inactive (softer) springs, thus deflecting the peripheral shape of the robot. The resulting movement exhibited flexible behaviour and amoeboid movement.

This paper is also motivated by the complex sensory and foraging behaviour of *P. polycephalum*, however it takes a different approach from that of Umedachi et al. Instead of trying to build an amoeboid robot with pre-existing oscillatory components, we investigate how oscillatory behaviour may *emerge* from the local interactions between simple component parts to generate self-organized amoeboid movement. The method uses a swarm based, or multi-agent, population exploiting self-organization to behave as a collective virtual material. We demonstrate the emergence of the oscillatory phenomena within the material, its utilisation for collective, amorphous and controllable amoeboid movement, and mechanisms for external control of the material by simulated chemoattraction and repulsion by light hazards. We conclude by discussing some novel properties of the approach within the context of existing research into swarm robotics and modular robotics.

## Methods:

To investigate the use of emergent oscillatory phenomena for collective movement we employ a simple extension to the particle approximation of *P. polycephalum* network adaptation in (Jones, 2010b) which was shown to generate dynamical emergent transport networks. In this approach a plasmodium is composed of a population of mobile particles with very simple behaviours, residing within a 2D diffusive environment. A discrete 2D lattice (where the features of the environment arena are mapped to greyscale values in a 2D image) stores particle positions and also the concentration of a local factor which we refer to generically as chemoattractant. The 'chemoattractant' factor actually represents the hypothetical flux of sol within the plasmodium. Free particle movement represents the sol phase of the plasmodium. Particle positions represent the fixed gel structure (i.e. global pattern) of the plasmodium. Particles act independently and iteration of the particle population is performed randomly to avoid introducing any artifacts from sequential ordering. Particle behaviour is divided into two distinct stages, the sensory stage and the motor stage. In the sensory stage, the particles sample their local environment using three forward biased sensors whose angle from the forward position (the sensor angle parameter, SA), and distance (sensor offset, SO) may be parametrically adjusted (Figure 1a). The offset sensors represent the overlapping and intertwining filaments within the transport networks and plasmodium, generating local coupling of sensory inputs and movement to form networks of particles (Figure 1c) and sheets of particles (Figure 1d). The SO distance is measured in pixels and a minimum distance of 3 pixels is required for strong local coupling to occur. The coupling effect increases as SO increases.

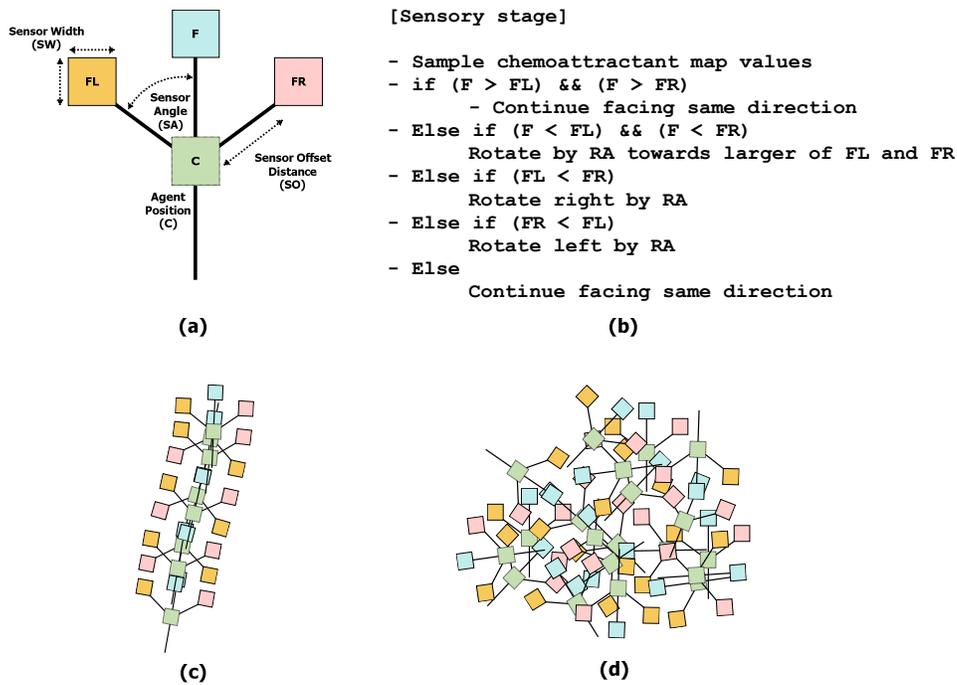

**Figure 1: Agent particle morphology, algorithm, and aggregation types.**

a) Schematic illustration of agent particle showing central position and offset sensors.
b) Pseudocode of agent sensory algorithm to align with strongest source from the forward biased sensors.
c) Schematic illustration of agent particles in transport network configuration
d) Schematic illustration of agent particles within simulated plasmodial sheet.

During the sensory stage each particle changes its orientation to rotate (via the parameter rotation angle, RA) towards the strongest local source of chemoattractant (Figure 1b). After the sensory stage, each particle executes the motor stage and attempts to move forwards in its current orientation (an angle from 0—360˚) by a single pixel. Each lattice site may only store a single particle and—critically—particles deposit chemoattractant into the lattice only in the event of a successful forwards movement (Figure 2a). If the next chosen site is already occupied by another particle the default (non-oscillatory) behaviour is to abandon the move, remain in the current position, and select a new random direction (Figure 2b).

Diffusion of the collective chemoattractant signal is achieved via a simple 3x3 mean filter kernel with a damping parameter (set to 0.07) to limit the diffusion distance of the chemoattractant. The low level particle interactions result in complex pattern formation. The population spontaneously forms dynamic transport networks showing complex evolution and quasi-physical emergent properties, including closure of network lacunae, apparent surface tension effects and network minimisation. An exploration of the possible patterning parameterisation was presented in (Jones, 2010a). Although the particle model is able to reproduce many of the network based behaviours seen in the *P. polycephalum* plasmodium such as spontaneous network formation, shuttle streaming and network minimisation, the default motor behaviour does not exhibit oscillatory phenomena and inertial surging movement, as seen in the organism. This is because the default action when a particle is blocked (i.e. when the chosen site is already occupied) is to randomly select a new orientation, resulting in very fluid network evolution resembling the relaxation evolution of soap films and the lipid nanotube networks seen in (Lobovkina et al., 2008). The oscillatory phenomena seen in the plasmodium are thought to be linked to the spontaneous assembly/disassembly of actomyosin and cytoskeletal filament structures within the plasmodium which generate contractile forces on the protoplasm within the plasmodium. The resulting shifts between gel and sol phases prevent (gel phase) and promote (sol phase) cytoplasmic streaming within the plasmodium. To mimic this behaviour in the particle model requires only a simple change to the motor stage. Instead of randomly selecting a new direction if a move forward is blocked, the particle increments separate internal co-ordinates until the nearest cell directly in front of the particle is free. When a cell becomes free, the particle occupies this new cell and deposits chemoattractant into the lattice (Figure 2c).

The effect of this behaviour is to remove the fluidity of the default movement of the population. The result is a surging, inertial pattern of movement dependent on population density (the population density specifies the initial amount of free movement within the population). The strength of the inertial effect can be damped by a parameter (pID) which determines the probability of a particle resetting its internal position coordinates, lower values providing stronger inertial movement. When this simple change in motor behaviour is initiated surging movements are seen and oscillatory domains of chemoattractant flux spontaneously appear within the virtual plasmodium showing characteristic behaviours: temporary blockages of particles (gel phase) collapse into sudden localised movement (solation) and vice versa. The oscillatory domains themselves undergo complex evolution including competition, phase changes and entrainment. We utilise these dynamics below to investigate the possibility of generating useful patterns of regular oscillations which may be coupled to provide motive force.

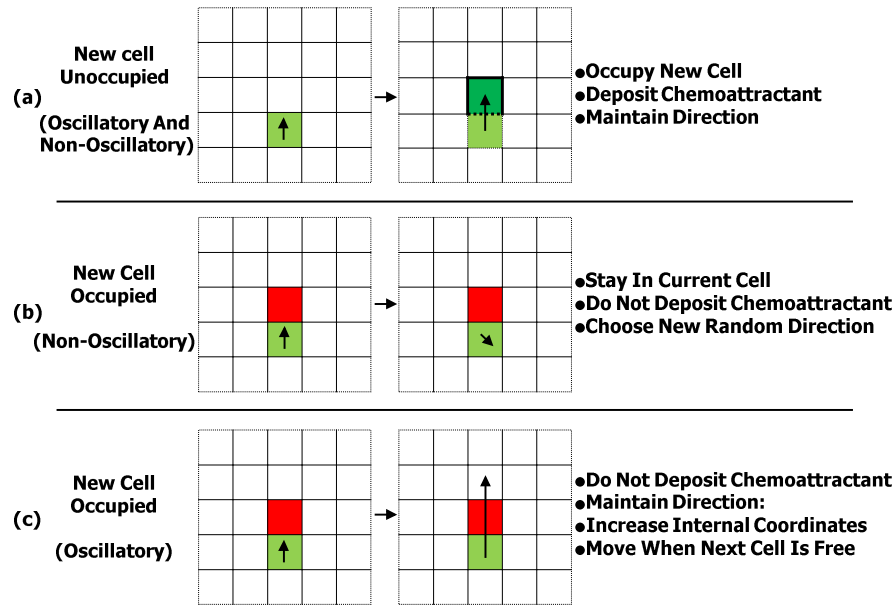

**Figure 2: Particle motor behaviour in non-oscillatory and oscillatory conditions**

a) A particle moves forward to occupy a vacant cell and deposits chemoattractant at the new location.
b) A non-oscillatory particle will select a new random location if the preferred cell is occupied.
c) If the preferred cell is occupied, the oscillatory particle will maintain the direction until a cell in front is free.

## Model Setup

The emergence of oscillatory behaviour in the model corresponds to differences in distribution of protoplasm within the plasmodium and subsequent changes in thickness of the plasmodium. In the *P. polycephalum* plasmodium, changes in thickness of the plasmodium membrane are used to provide impetus (pumping of material through the vein network, or bulk movement of the plasmodium). There is known to be a reciprocal relationship between the spontaneous contraction of the plasmodium and the subsequent transport of protoplasm away from that region (Takamatsu et al., 2000). Thus the region undergoing contraction becomes thinner (allowing more light transmission when illuminated) as protoplasm is being transported away from the contraction zone. Protoplasmic tubes connected to the contraction zone become thicker as more protoplasm is transported from the zone (thus allowing less light transmission). In the computational model the transport of particles represents the free flux of protoplasm within the material and a local increase in flux (particle movement) is indicated in the supplementary video recordings by a temporary increase in greyscale brightness at this location. A decrease in the bulk movement of particles represents local congestion (lack of transport) and is indicated by a decrease in greyscale brightness (since deposition of chemoattractant factor only occurs in the event of successful forward movement). Supplementary video recordings showing the initiation and entrainment of oscillatory phenomena, and recordings from specific figures in the paper text can be found at: http://uncomp.uwe.ac.uk/jeff/collectivemovement.htm. For visual clarity in the static images, the greyscale images are inverted (dark areas indicate greater flux).

The particle population environment is a 2D lattice, represented by a digitised image configured to represent the habitat of the experimental plasmodium. The habitat is composed of 'wall' regions where particle occupancy and movement cannot occur, 'vacant' regions where occupancy and movement are possible and, where relevant, 'stimulus' regions which provide attraction stimuli, or repulsion stimuli, to the particle population. In the experiments there is an initial period where oscillatory behaviour is not initially activated and this results in self-organized regularly spaced domains within the collective. When oscillatory motor behaviour is induced these regular domains collapse and small domains of disorganized oscillators begin to emerge. The oscillation domains are composed of local areas of flux (brighter) and obstruction (darker). Over time these domains compete and coalesce, causing entrainment of the population into regular oscillation patterns, influenced by both the particle sensory parameters and also by the shape of the experimental arenas.

## Emergence of Amoeboid Movement

P. polycephalum utilises internal protoplasmic transport to migrate towards nutrient sources and away from hazardous sources (Durham and Ridgway, 1976), and adapts its gross body plan to changing environments (Nakagaki et al., 2004; Nakagaki et al., 2007). Small plasmodia can shift the entire plasmodium away from unfavourable conditions such as bacterial or fungal contamination. The plasmodium is also notable for its ability to survive physical damage; fragments of plasmodium excised can survive independently and individual plasmodia may be fused to form a single organism. To utilise a vehicular analogy, *P. polycephalum* not only represents the internal mechanicals (motive force mechanism, transmission coupling), but also the moving vehicle itself, and is a vehicle which can survive the removal of parts, the introduction of new foreign parts and the repair of damaged parts.

We set out to explore the behaviour of the particle collective to assess its behaviour when compared to a fragment of *P. polycephalum* plasmodium. When oscillatory motor behaviour is not used the particle collective condenses into a uniform circular shape as the initial transport network condenses (Figure 3a). The non-oscillatory blob shows regular vacancy domains (dark areas) and the fluid particle motion afforded by the non-oscillatory motor condition ensures that the blob is cohesive and takes a minimal shape. The non-oscillatory blob is also resilient to external perturbation. When excited by an externally applied source of chemoattractant (Figure 3a, mouse position in fourth image), the deformation of the collective induced by the stimulus as it is attracted to the stimuli is repaired when the stimulus is removed, the collective returning to its minimal shape.

When oscillatory motor behaviour is initialised the regular domains collapse as the particle motion becomes less fluid (Figure 3b) and oscillations travel through the collective. Because the small collective is not constrained by any externally applied pattern the oscillations distort the shape of the collective. When the pID parameter is further reduced to 0.01 there is even greater restriction on the fluidity of individual particle movement and the oscillations become stronger and distort the collective's boundary significantly. The large shift of a mass of particles causes the collective to move across its environment. The cohesion of the collective is maintained but other SA/RA parameter settings, combined with lower sensor interaction (SO) distance, can result in the fragmentation of the collective (see supplementary material for examples of oscillation patterns using different SA/RA combinations).

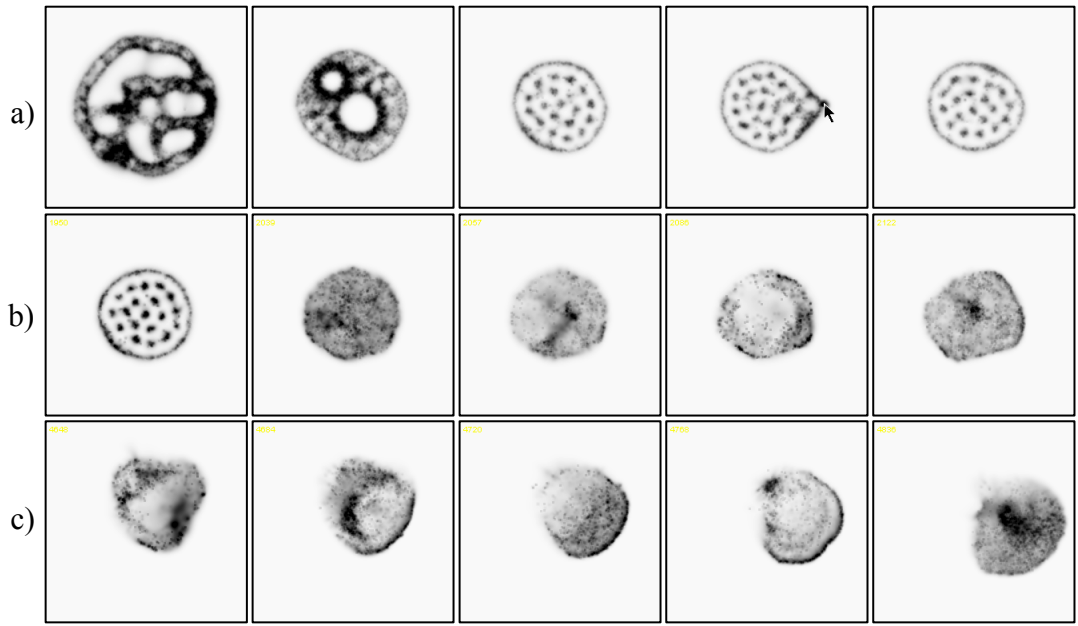

**Figure 3: Initiation of oscillatory behaviour and amoeboid movement in simulated plasmodium**

Images show flux within simulated plasmodium over time. Images anchored to fixed location.
a) Condensation of blob material in non-oscillatory behaviour shows vacancy domains and resilience to deformation. Collective of 9380 particles, SA90, RA45, SO15
b) Initial non-oscillatory collective with regular vacancy domains (left) and onset of oscillatory behaviour at 1950, 2039, 2057, 2086 and 2122 scheduler steps.  pID set to 0.05. All other parameters as in (a)
c) Reduction of pID to 0.01 results in stronger oscillations and amoeboid movement at 4648, 4684. 4720, 4768 and 4836 scheduler steps. All other parameters as in (a)

## Persistent Movement in a Small Blob Fragment

The amoeboid movement seen in Figure 3 occurs because the oscillation waves distort the boundary of the collective whilst it is still able to maintain a cohesive whole. Because the population maintains its cohesion, any distortion of the boundary on one side must result in a shift in population distribution from the opposite side (since the collective is non-compressible and occupies a fixed area). The diameter of the collective (a function of the number of pixel sized particles comprising it) must be large enough to for oscillations to emerge and to confine an oscillation pattern within it.

When the collective is comprised of only a relatively small number of particles, the distortion of the boundary forms an approximately semicircular domed shape and the small number of particles ensures that the collective cannot maintain a fully circular shape. However the persistence of forward movement generated by the oscillatory motor behaviour at low pID values causes the dome shape itself to be maintained over time, and the small blob fragment is able to move forwards (Figure 4a). Movement of the fragment is relatively smooth and different from the pulsatile motion observed in larger collectives. Particles move towards the front of the domed profile (Figure 4b, dark pixels) and then, over time, move to the side. Particles at the sides of the fragment ultimately fall behind only to re-enter the dome at the centre. Higher pID values result in more frequent changes of direction of the fragment as the dome shaped front profile cannot be maintained. If the population size is increased, the single sided dome shape cannot be maintained and the resultant motion becomes pulsatile and chaotic. The movement of small blob fragments in the particle collective mimics amoeboid motion observed in *P. polycephalum* on a non-nutrient substrate (Figure 4c) which has been shown to be equivalent to the propagation of wave fragments in sub-excitable media (Adamatzky et al., 2008; Adamatzky, 2010b). Both artificial and real blobs exhibit reflection (reversion of direction) when encountering the boundary of their environment and can both be directionally guided by the placement of attractants and repellents (Adamatzky, 2010a).

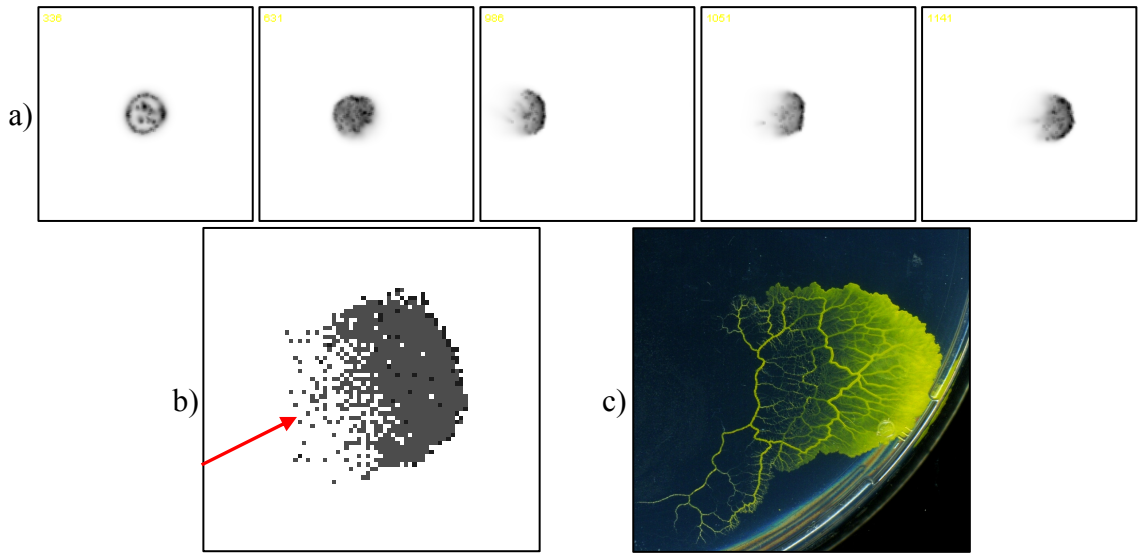

**Figure 4: Persistent forward movement of blob fragments in small simulated plasmodia**

Images show flux within simulated plasmodium fragments over time. Images anchored to fixed location.
a) Non-oscillatory condition, initiation of oscillatory behaviour and (final 3 images) self perpetuating transport of blob fragment. Chemoattractant flux concentration is greatest at the front of the dome shape
b) Enlargement showing particle composition of the moving blob fragment. The persistent shape is maintained despite repeated turnover of component parts. Population size 900 particles, SA90, RA45, SO9, pID 0.001
c) Independently moving small plasmodial fragment of *Physarum polycephalum*

## External Influence of Collective Movement – Attraction and Repulsion

Movement of the *P. polycephalum* plasmodium is strongly affected by local environmental conditions. Attractant sources (such as increasing temperature gradients and chemoattractant nutrients) cause the plasmodium to move and grow towards the attractants whereas repulsive sources (salts, dry regions) cause the plasmodium to try to avoid such regions (Durham and Ridgway, 1976). The plasmodium is able to integrate many separate localised inputs to compute its response to the environment. One method in which this is achieved is by the modulation of local oscillation patterns in response to attractants or hazards – attractants tend to increase localised oscillation strength and hazards decrease oscillation strength. We set out to see if a localised response to external influences could be used to govern the collective movement of the particle population.

Attractant sources were previously used as a method to confine the collective to a region by pinning it down. By externally presenting an attractant source (effectively a simulated nutrient source) near to a cohesive blob of virtual material (circle in Figure 5a) a concentration gradient emerged from the source (Figure 5b). When the diffusion gradient reached the sensors of the closest particles at the front of the collective it provoked local movement towards the source. The cohesion within the collective resulted in a pseudopodium-like extension of the border region which extended towards the source (Figure 5c), ultimately engulfing it. Travelling waves spontaneously emerged within the collective which were directed at the source, causing the collective to shift its position towards the source (Figure 5d). Consumption of the source was simulated by simply decrementing the value projected to the diffusion field when the source was covered by a particle. When the source was consumed by the population, the collective regained its previous approximately circular shape.

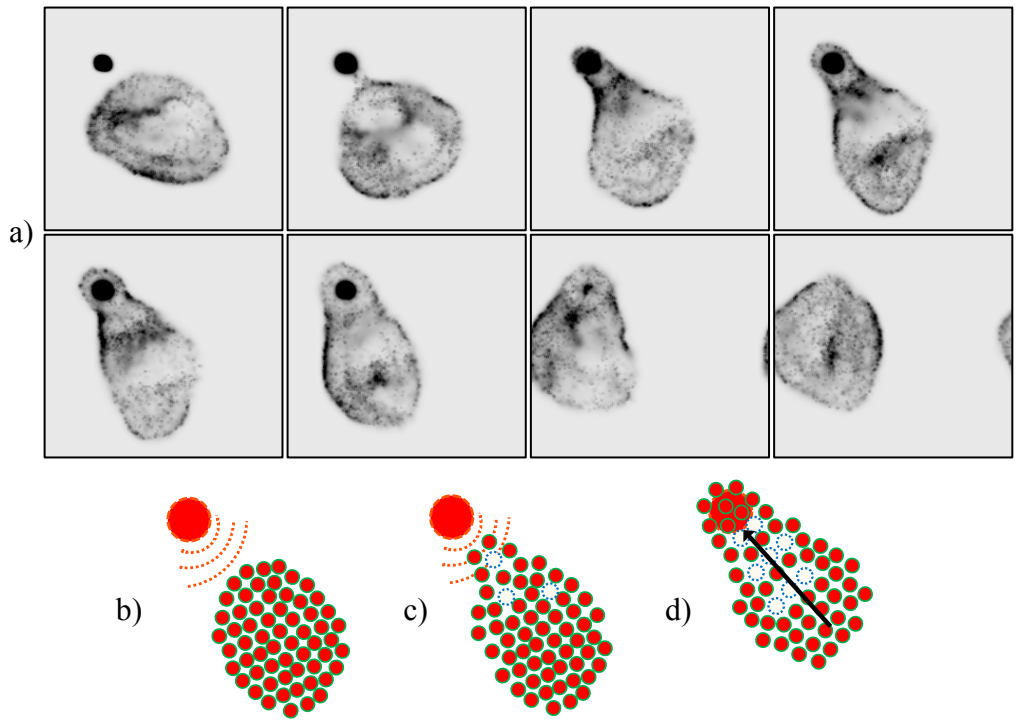

**Figure 5: Control of amoeboid movement by attraction**

Images show flux within simulated plasmodium over time. Images anchored to fixed location.
a) A persistent source of chemoattractant is placed into the diffusion field which provokes the extension of the collective. The collective engulfs the source by moving via travelling waves towards the source. When the source is exhausted the collective re-adopts its original approximately circular shape.
b) Schematic illustration of attraction of the collective by the nutrient source
(c) Migration of leading particles towards the source
(d) Emergence of travelling waves pulling the collective to engulf the source.

To approximate the repulsion of the collective to hazardous sources such as the simulated response to irradiation by visible light we added a condition to the sensory stage of the algorithm to the effect that if any particles of the collective were in a region exposed to 'light' (a defined area within the arena), those particles would have their sensitivity to chemoattractant diminished whilst they remained in this region (this achieved by multiplying the sampled sensor values with a weighting factor less than 1, lower values generating a stronger response to irradiation). The effect of exposing regions of the collective to simulated light damage was that the collective immediately started to move away from the irradiated region (Figure 6a). Specifically, oscillation waves moved from the irradiated region towards the unexposed regions. The shift of particles from the irradiated region eventually moved the collective away from the stimulus. The cause of movement away from the light can be found at the interface between irradiated and unexposed areas. Before irradiation (Figure 6b) all regions of the collective are equally attractive to the particles (subject to fluctuations caused by discrepancies in particle movement and intrinsic oscillations within the collective). There is a strong coupling between the particles in the collective caused by the offset sensor distance. Some of the particles at the interface of the irradiated region will receive input from the unexposed region and will be attracted to that area because the chemoattractant concentration in unexposed areas is perceived as greater due to the damping in irradiated regions (Figure 6c). The movement of particles near the interface towards unexposed regions causes both new vacant spaces (Figure 6d) and also an increase in chemoattractant concentration (because only mobile particles deposit chemoattractant). This results in a further increased attraction to the interface region, until eventually the entire collective has migrated from the irradiated region.

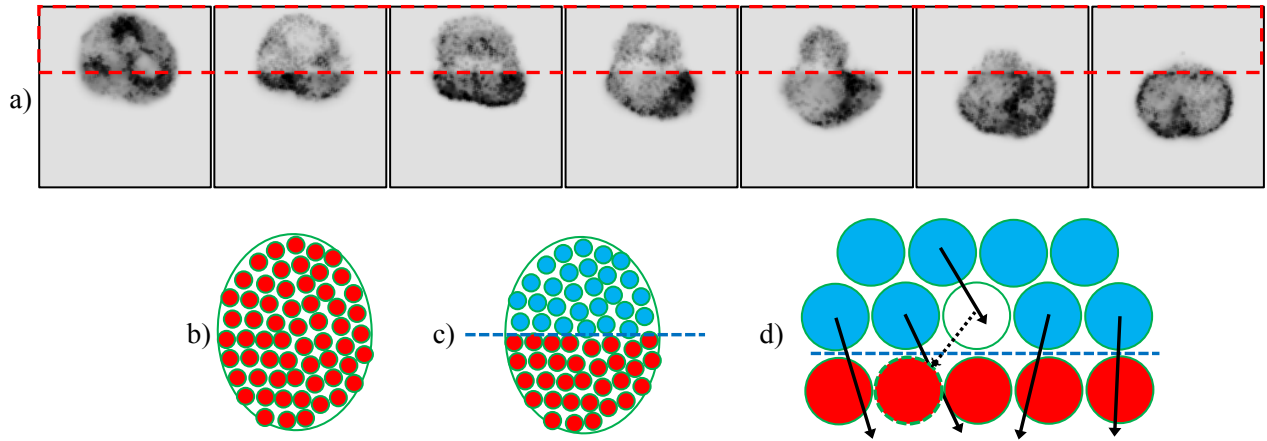

**Figure 6: Avoidance of simulated light irradiation by particle collective**

Images show flux within simulated plasmodium over time. Images anchored to fixed location.
a) Area within dashed box is stimulated with simulated light irradiation. Particle collective oscillates sending travelling waves towards the region which is unexposed, moving the collective away from the irradiated region.
b) Schematic illustration of condition before irradiation – all particles sense equal concentration of chemoattractant
c) Irradiated areas (top) are perceived as weaker concentration
d) Particles at the irradiation interface are more attracted to unexposed areas. Migration across interface causes chemoattractant deposition, causing further attraction to region and vacant space.

## Morphological Adaptation of the Collective

The previous results demonstrated that the collective changes its shape during self-oscillatory behaviour and also in response to simulated attractants and hazards. The collective retains its typically circular shape due to the cohesion of the individual particles making up the population. When the morphology of the collective is disturbed by its movement towards, or away from, externally applied stimuli it can reform the original shape when the stimulus is removed. An adaptive morphology is a very desirable property in robotic devices since it imbues the robot with great flexibility of size and movement, enabling it to traverse environments which traditionally are difficult to navigate (for example narrow spaces, gratings etc.). This feature is only possible because the properties of the movement and guidance of the collective are distributed throughout the collective and not located in fixed sized and inflexible units as is the case with conventional robotic systems. This is also the case with the *P. polycephalum* plasmodium which adapts its shape and growth patterns in response to its environment. One of the most remarkable properties of the plasmodium is the ability to survive external damage beyond simple attraction and repulsion. A piece of plasmodium excised from the growing tip can survive, and indeed continue to move and grow as an independent entity. Furthermore two independent plasmodia can fuse to form a single plasmodium when placed in close proximity. These phenomena are not only desirable from a robotics perspective in terms of resilience and damage repair, but offer new and as yet little explored opportunities in robotic movement and control.

We set out to find if the particle collective could also replicate these highly desirable features as seen in the real plasmodium. We took a large single oscillating collective (5000 particles) and applied a narrow band of hazardous simulated light irradiation through its centre (Figure 7a, dashed box represents irradiated area). Particles immediately began to surge away from the irradiated region on both sides and the collective narrowed in diameter and became further pinched in shape until the collective was cleaved into two independently controllable blobs. The cleavage mechanism can be applied in different ways. For example both resulting blobs can be of equal size and have similar oscillatory properties. Alternately it is possible to cleave the collective in such a way to have one large stationary blob and one smaller mobile blob (recall that a blob which is small enough will be able to move spontaneously in a persistent direction).

It is possible to guide each blob independently using either a *pulling* type mechanism (externally applied attractants) or a *pushing* type mechanism (simulated irradiation). In Figure 7b we guide the lower right blob (arrowed) towards the larger blob by pushing it from its opposite side with simulated irradiated light. As the blobs become closer (specifically, to a separation distance which is sufficient for the border particles in each blob to sense the chemoattractant flux in the other blob) the closest border regions of each blob surge towards each other and a single larger collective is formed by the fusion. Cleavage and guidance of the particle collective by simulated hazards reproduces the control of *P. polycephalum* plasmodial migration (Figure 7c) by repellents (Adamatzky, 2010a).

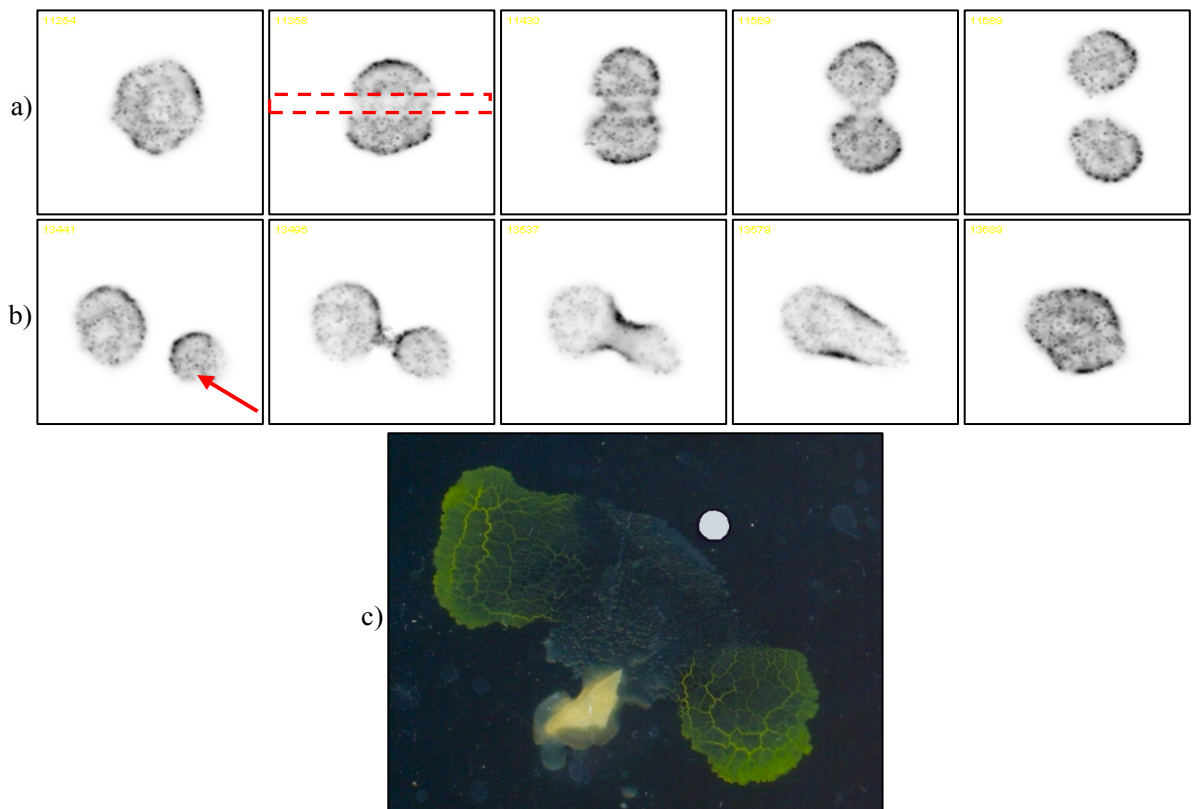

**Figure 7: Controlling blob morphology by splitting and fusion.**

Images show flux within simulated plasmodium over time. Images anchored to fixed location.
a) 'Blob' of aggregate particles is split by applying simulated light irradiation (dashed box), disrupting particle flux in that region. The single blob separates into two smaller blobs, each capable of individual oscillation and external control.
b) Fusion of two independent particle aggregates. The blob on the lower right is guided diagonally upwards in the direction of the arrow towards the larger blob. The two independent blobs fuse forming a single aggregate.
c) Cleavage of plasmodium of *P. polycephalum*. Fragment of plasmodium initialised at triangular region. Crystal of potassium chloride placed at circular region. Diffusion of potassium chloride cleaves the plasmodium into two independent fragments.

As an example of the robotic flexibility endowed by the adaptive morphology of the collective, and the guidance mechanisms enabled by its external control, we show in Figure 8 three examples of how a blob can be guided externally (in this case a repulsive *pushing* by simulated irradiation) in order to traverse narrow, separated or tortuous paths where the path may be much narrower than the diameter of the collective itself. In Figure 8a, the blob has a diameter of approximately 52 pixels and is guided through a channel of width 30 pixels by automatically reconfiguring its morphology. In Figure 8b, the blob is split into two parts by irradiating a rectangular region at the entrance of the channel. The two separated blobs may be guided separately by simulated radiation and are re-fused near the exit of the channels. Finally, in Figure 8c, the blob automatically separates its structure in response to the different obstacles and re-forms its shape by mutual attraction and cohesion of the particles. No fine control of the individual components, nor complex pre-determinism of path choice, is necessary; the collective is guided by simple avoidance of the simulated irradiation (the irradiation location is not shown, but follows the previous examples of simply exposing the rearward part of the collective to push it forwards). When clear of the obstacles, the original circular shape is reformed. If the blob is returned backwards through the obstacle path again the path chosen can be somewhat different to the original path with the same result.

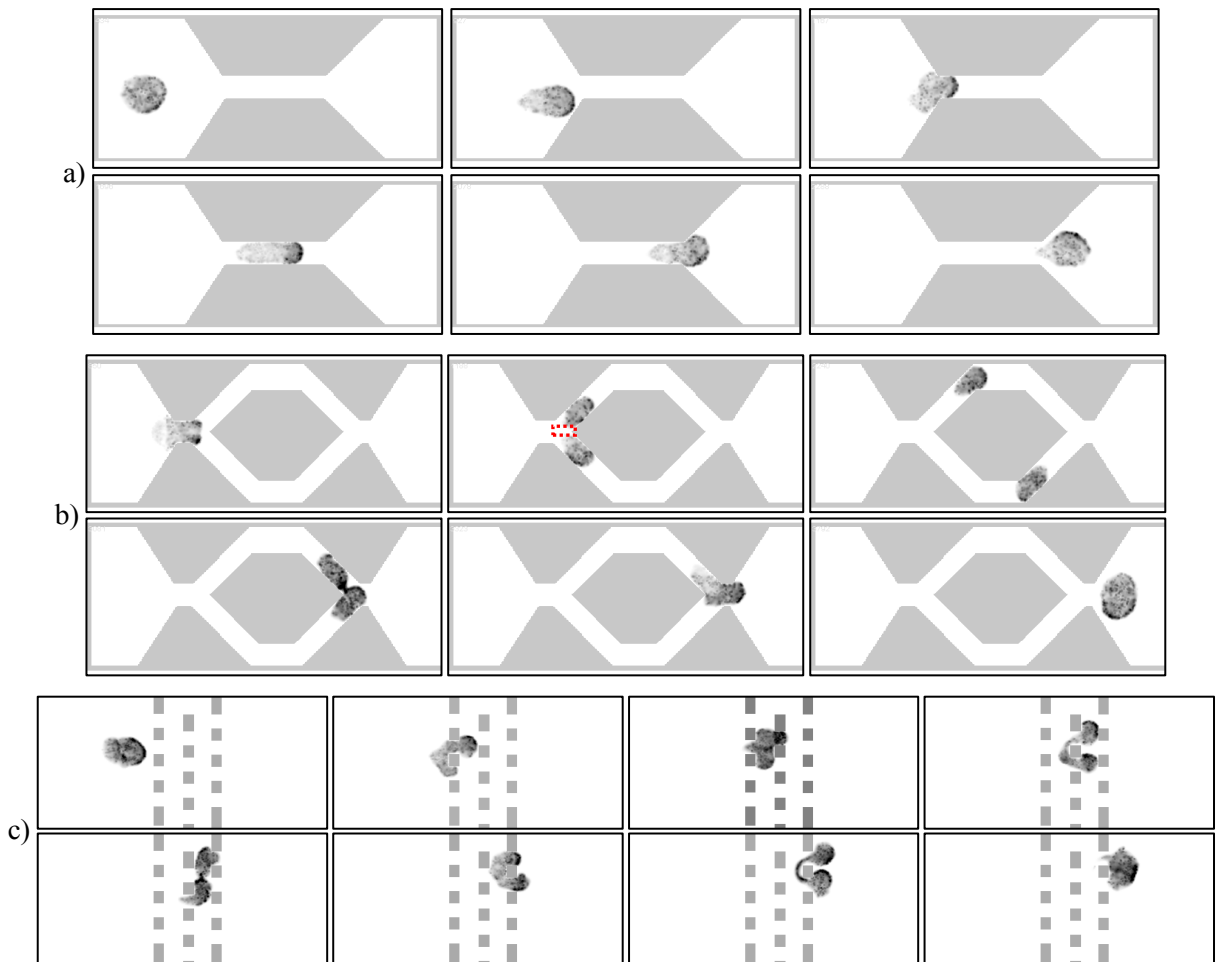

**Figure 8: Exploiting the adaptive morphology of the collective.**

Aggregate blob of simulated plasmodium is *pushed* by the application of external simulated light irradiation (illuminated at the left side of the aggregate, not shown).
a) Large aggregate moves through a channel narrower than the aggregate diameter
b) Aggregate is cleaved by hazard irradiation (boxed) and guided separately before being refused into original size
c) Aggregate is pushed through tortuous narrow grating and reforms original aggregate shape.

# Conclusion and Discussion:

We have examined the problem of generating collective amoeboid movement in a multi-agent approximation of the plasmodium of true slime mould *Physarum polycephalum* using self-organization between simple non-oscillatory component parts. *P. polycephalum* is attractive because it satisfies many physical and computational properties which are desirable in robotics applications (self-oscillatory, simple components, distributed sensory and motor control, integration of multiple sensory stimuli, amorphous and adaptive shape, amenable to external influence, resilience to damage, self repair) and can be regarded as a living example of a so-called smart material because of the way in which it combines robotic and control functions in a distributed manner throughout its constituent material.

How can such complex behaviour emerge from the interactions between simple component parts? Answering this question may provide insights into the development of non-living smart materials which would have the advantages demonstrated by the *P. polycephalum* plasmodium, but without some of its limitations, such as slow speed, fragility and unpredictability. Because of its simple components and structure, any search for a 'secret source' of the plasmodium's complexity would be fruitless. Instead we set about answering the question by posing in reverse: rather than try to find out how the organism produces such complex behaviour from simple parts, is it possible to artificially generate similarly complex behaviour from simple parts and interactions? We utilised a previous particle model of emergent transport networks where a collective of identical particles with identical behaviour was used to construct and minimise synthetic and emergent transport networks. We modified its motor algorithm in a very simple way so that, instead of smooth network flow, we obtained a more resistant flow of particles. The addition of transient resistance to particle flux was sufficient to generate complex oscillatory behaviour strongly similar to that observed in the plasmodium itself.

The condensation of the collective in its virtual environment resulted in cohesive blob-like sheets of virtual material held together by cohesion which exhibited spontaneous internal oscillations. The oscillations were composed of regular spatial domains of mobile particles and temporarily restricted particles. The wave-like propagation of the oscillatory domains resulted in a collective amoeboid movement as the perimeter of the particle collective was deformed by the internal oscillations. External control of the blob sheets was achieved by stimulating the collective with simulated attractants (*pulling* the collective towards the source) and repellents ('pushing' the collective away from simulated light irradiation). We were also able to reproduce the resilience of the *P. polycephalum* plasmodium to damage by cleaving the collective into two separate and independent blobs and fusing two independent blobs to form a single functional blob. Finally the adaptive morphology of the collective was demonstrated by guiding the collective through an obstacle field narrower than the diameter of the collective itself. The flexible modularity and morphology seen in this approach builds upon previous work on the external influence and control of robotic devices (Bojinov et al., 2002; Salemi et al., 2001) by enabling separate and functional modules of arbitrary size to be created and re-formed from a single collective.

The results demonstrate how very simple and local low-level interactions within a simple material can generate complex and emergent behaviour which appear to transcend the capabilities of the simple matter of which they are composed. As noted by previous swarm approaches, there need not be anything magical or special about the properties of these materials; the complexity emerges merely from their interactions (Buhl et al., 2006; Reynolds, 1987; Vicsek et al., 1995). This complex behaviour is harnessed effortlessly by organisms such as *P. polycephalum* as part of a parsimonious survival strategy, enabling their persistence in unpredictable, changeable and hazardous environmental conditions. By understanding the generative mechanisms underlying the complex behaviour it may be possible to incorporate these features within real physical materials for small scale robotic devices. By utilising the robotic substrate material itself for distributed computation, transport and movement it may be possible to reduce the total number of component parts and also reduce the number of different types of components, thus further simplifying the production of the devices.

## Acknowledgements:

The work was partially supported by the Leverhulme Trust research grant F/00577/1 "Mould intelligence: biological amorphous robots".